\documentclass{article}
\usepackage{spconf,amsmath,graphicx,hyperref}

\usepackage{algorithm}%
\usepackage{algorithmicx}%
\usepackage{algpseudocode}%
\usepackage{listings}%
\usepackage{amssymb}
\usepackage{color}
\usepackage{tabularray}
\usepackage{todonotes}

\usepackage{graphicx}
\usepackage{array}
\usepackage{caption}
\usepackage{kantlipsum}
\usepackage{multirow}

\title{Regularizing INR with Diffusion Prior for Self-Supervised 3D Reconstruction of Neutron Computed Tomography Data}
\name{
Maliha Hossain,
Haley Duba-Sullivan,
Amirkoushyar Ziabari
\thanks{Corresponding author's email address: ziabariak@ornl.gov. 
This manuscript has been authored by UT-Battelle, LLC, under contract DE-AC05-00OR22725 with the US Department of Energy (DOE). The US government retains and the publisher, by accepting the article for publication, acknowledges that the US government retains a nonexclusive, paid-up, irrevocable, worldwide license to publish or reproduce the published form of this manuscript, or allow others to do so, for US government purposes. DOE will provide public access to these results of federally sponsored research in accordance with the DOE Public Access Plan (https://www.energy.gov/doe-public-access-plan).}}
\address{Oak Ridge National Laboratory, 1 Bethel Valley Road, Oak Ridge, TN, USA 37830}
\begin{document}
%
\maketitle
\begin{abstract}

Recently, generative diffusion priors have made huge strides as inverse problem solvers, including the ability to be adapted for inference on out-of-distribution data. Concurrently, implicit neural representations (INRs) have emerged as fast and lightweight inverse imaging solvers that are amenable to hybrid approaches that combine learned priors with traditional inverse problem formulations. 
In this paper, we present a diffusive computed tomography (CT) inversion framework for regularizing INRs called Diffusive INR (DINR), designed to enable high-quality reconstruction from sparse-view neutron CT. Pretrained purely on synthetic data, DINR is evaluated on simulated and experimentally obtained observations of concrete microstructures, where traditional reconstruction methods suffer substantial degradation when the number of views is reduced. Our approach delivers superior performance, reduces reconstruction artifacts, and achieves gains in PSNR and SSIM, enabling accurate micro-structural characterization even under extreme data limitations compared to state-of-the-art sparse-view reconstruction techniques. 

\end{abstract}



\section{Introduction}
\label{sec:intro}

Neutron Computed Tomography (CT) is an important imaging modality due to its unique ability to characterize volumes by hydrogen distribution~\cite{kardjilov2011neutron}.
This capability is critical in a number of settings, including manufacturing hydrogen fuel cells~\cite{ziesche2022high} and lithium-ion batteries~\cite{ziesche20204d},
studying water transport in plants and soil~\cite{totzke2017capturing,shafabakhsh20244d},
and safety monitoring of concrete components for radiation shielding and structural integrity \cite{anunike2024ultra,yan2020reconstruction}. 
Conventional CT reconstruction, generally using projection measurements with a few thousand views, is performed using analytical methods such as Filtered Back Projection (FBP).
Many CT applications necessitate sparse view sampling for fast acquisition, such as real-time surgical guidance, security screening, etc. 
Neutron CT, in particular, due to its low beam flux, requires much longer exposure times to achieve measurements with acceptable signal-to-noise ratio~\cite{wang2025fstu}.
However, FBP typically cannot handle view sparsity below Nyquist without generating severe artifacts~\cite{li2021sparse}, motivating the development of model-based iterative reconstruction (MBIR) that includes a prior model term in a cost function formulation. 
MBIR requires hand-constructed priors such as Total Variation (TV) and other Markov random fields of higher orders like qGGMRF~\cite{bouman1993generalized,venkatakrishnan2021algorithm}.
However, diffusion-based inverse problem solvers are able to model more complex image properties and are at the forefront of powerful generative priors for solving ill-posed inverse problems~\cite{barbano2023steerable,chung2024deep}.

Sparse-view CT is an ill-posed inverse problem that occurs when the  scanner collects fewer projection views than required by the Nyquist sampling theorem~\cite{li2021sparse}.
There have been several notable advances in unsupervised deep learning-based approaches for solving inverse imaging problems.

Implicit neural representations (INRs)~\cite{sitzmann2020implicit} have emerged as a powerful tool in the field of 3D imaging by learning to use neural networks to represent a volume with a continuous function.
While this in turn allows them to capture a reconstructed volume with a memory-efficient and resolution-independent representation, 
standard implicit neural representations exhibit a strong spectral bias toward low-frequency components, which can lead to poor reconstruction of high-frequency structure and unstable behavior when supervision is sparse or ill-conditioned \cite{tancik2020fourier}.
However, their ability to easily integrate learned priors and CT forward models makes them a powerful tool to be exploited in conjunction with other advanced prior modeling techniques ~\cite{zha2022naf, shi2024implicit}.

Steerable Conditional Diffusion (SCD) \cite{barbano2023steerable} is an effective approach for leveraging measurement data to guide a pretrained diffusion model towards a target distribution at test time.
SCD alternates between updating the diffusion model weights and producing an estimate for posterior mean.
In \cite{chung2024deep}, the authors propose DD3IP, a 3D Deep Diffusion Image Prior, which is a generalization of SCD that can coherently reconstruct across orthogonal spatial dimensions in a 3D volume of out-of-distribution (OOD) parallel beam CT data using a UNet denoising diffusion model. 
One key contribution of DD3IP is the conclusion that posterior sampling in SCD can be performed agnostically of the choice of diffusion inverse problem solver (DIS), resulting in a modular framework where we can plug in state of the art CT measurement inversion models. 
In this work, we build on that conclusion as follows:
\begin{enumerate}
\item We formulate a regularized INR DIS within a DD3IP framework and demonstrate its ability to generate high-quality sparse-view parallel-beam neutron CT reconstructions. Our INR incorporates the denoised diffusion output into a proximal formulation of the INR loss function. 

\item Our software implementation improves on \cite{chung2024deep}'s by adopting a distance driven parallel beam projector from Tomosipo, a pythonic wrapper for the ASTRA-toolbox to enforce data consistency, and for better modularity. 
Our code and data will be made available online. 

\item We present a comparison against MBIR with a qGGMRF prior, which is a more realistic model than the more commonly presented TV prior. DINR can perform on par with qGGMRF in sparse-view reconstruction and even outperform it in the ultra-sparse regime.
\end{enumerate}


\section{Background}
\label{background}

\subsection{Implicit Neural Representations (INRs)}  
INRs model images or volumes as continuous functions parameterized by a neural network, typically a multilayer perceptron that maps spatial coordinates to signal values, e.g., attenuation coefficients~\cite{park2019deepsdf, sitzmann2020implicit}. 
A widely used INR architecture is the sinusoidal representation network (SIREN)~\cite{sitzmann2020implicit}, which employs periodic sinusoidal activation functions in each layer of an MLP to better represent high-frequency content. 
The coordinate-based formulation of INRs enables resolution-independent reconstruction, differentiable sampling, and compact encoding of entire volumes within a small set of network weights. 
INRs can be optimized directly against measured CT projections and flexibly integrated with physics-based forward models and learned priors, making them attractive for model-based or plug-and-play reconstruction. 
Recent works have demonstrated INR-based approaches for sparse-view and limited-angle CT, leveraging implicit functions to jointly learn geometry and attenuation while regularizing through learned priors or deep image prior~\cite{zha2022naf, shi2024implicit}. 
Extensions to this framework have incorporated additional input channels, such as FDK reconstructions, to provide an initial estimate that accelerates convergence and improves reconstruction quality~\cite{zhou2025rho}. 
However, INR-based reconstructions can be slow to optimize and may lack strong image priors unless combined with architectural regularization (e.g., dropout) 
or learned denoisers~\cite{rahaman2019spectral, ramasinghe2023much}.
\vspace{-0.3cm}

\subsection{3D Deep Diffusion Image Prior (DD3IP)}
A standard denoising diffusion probabilistic model (DDPM) \cite{ho2020denoising} consists of a forward process, where Gaussian noise is gradually added to a sample, and a reverse process, where a neural network is trained to denoise each step.
Denoising diffusion implicit models (DDIMs)~\cite{song2020denoising} were introduced as an efficient and deterministic variant of standard diffusion models.
SCD can "steer" or guide a pretrained inverse imaging diffusion prior to adapt to an OOD inference sample at test-time.
Instead of using the fixed DDIM reverse process, SCD applies gradient-based updates to guide the sample toward OOD data.
\cite{chung2024deep} showed that SCD is a special case of Deep Image Prior and formulated a coherent multi-slice reconstruction framework for parallel beam CT that alternates between adapting the diffusion model weights for OOD inference data and sampling from the posterior using DDIM.  
They further showed that the resulting DD3IP framework is agnostic of the strategy to estimate the posterior mean in the DDIM step as long as it is informed by the weight-adapted diffusion model.

\vspace{-0.3cm}


\section{Method}
\label{sec:method}

Let $y=Ax+n$, where $x$ is a 3D tensor of attenuation coefficient values mapped over a volumetric lattice $S$, $y$ is its corresponding projection, $A$ is a parallel beam CT projection matrix, and $n$ is additive noise approximating the actual noise in the measurement. 
The analytical solution of this problem $x=A^*y$ can be found using FBP. 

SCD can be presented in a deep image prior framework as
\begin{align}
    \mbox{for }t=T\dots 1:\theta_{t-1} &\gets\arg\min_{\theta_t}\mbox{MSE}\left(y, AD_{\theta_t}(x_t|y)\right)  \label{eq:adapt} \\
    x_{t-1} &\gets DDIM_{\theta_{t-1}}\left( D_{\theta_{t-1}}(x_t|y),\eta\right) \label{eq:post_sample}
\end{align}
where $t\in[0,T]$ is the diffusion time-step variable, $\theta_t$ is the diffusion model weights at time $t$, $D_{\theta_t}$ is the denoising diffusion model, and $x_t$ is the estimate at time $t$. Additionally, $\eta\in[0,1]$ is a random variable that controls stochasticity and MSE is the mean squared error loss. 
DDIM sampling is given by 
\begin{equation}\label{eq:ddim}
\begin{aligned}
    DDIM_{\theta}(\hat{x}_t,\eta):= &\sqrt{\alpha_{t-1}} \hat{x}_t \\
    &+ \sqrt{1-\alpha_{t-1}}\left(\eta\epsilon+(1-\eta)\epsilon^{\theta}\right),
\end{aligned}
\end{equation}
where $\alpha_{1:T}\in(0,1]^T$ is a decreasing sequence that parameterizes the reverse diffusion process \cite{song2020denoising}, $\epsilon$ is spherically interpolated Gaussian noise \cite{chung2024deep}, and $\epsilon^{\theta}$ is the noise removed by $D_{\theta}$.

The INR model $F_{\phi}$ with associated weights $\phi$ is trained to map coordinates in the cubic lattice $S$ to attenuation coefficeint values. 
Our INR is formulated to also accept the FBP estimate, $A^*y$, as an input to further inform reconstruction as in~\cite{zha2022naf}. 
Since the INR is regularized by the diffusion model, we additionally include the current diffusion model estimate $\hat{x}_{t}=D_{\theta_t}(x_t|y)$ in the loss to be minimized. Altogether, we define a proximal loss for updating the INR weights:
\begin{equation} \label{eq:inr_loss}
\begin{aligned}
\mathcal{L}_{\phi}(S, y, \hat{x}_{0|t}, \rho) 
&= \text{MSE}\left(A F_\phi(S, A^*y),\, y\right) \\
&\quad + \rho\, \text{MSE}\left(\hat{x}_{t},\,   F_\phi(S, A^*y)\right),
\end{aligned}
\end{equation}
where $\rho$ is a user-defined parameter controlling the influence of the diffusion model estimate. When we initialize the INR weights, we use $\rho = 0$, which gives the standard MSE loss. Since $\hat{x}$ is not needed in this case, we use the notation $\mathcal{L}_{\phi_{t-1}}(S, y, -, \rho=0)$. The resulting DDIM sampling including the INR is given by 
$DDIM_{\theta_{t-1}}\left(F_{\phi_{t-1}}(S, A^*y), \eta\right)$
where $\phi_{t-1} =\arg\min_{\phi}\mathcal{L}_{\phi}(S, y, \hat{x}_{t}, \rho)$.

We replicate the initialization strategy from \cite{chung2024deep} where the diffusion model weights are pretrained on synthetic ellipsoid data and $x_T$ is initialized at the start of the reverse diffusion process as 
\begin{equation}\label{eq:xt_update}
    x_T \gets \sqrt{\alpha_T} A^*y + \sqrt{1-\alpha_T}\epsilon*w,
\end{equation}
where $A^*y$ represents the low frequency component of the solution corrupted with $\epsilon$. 
We additionally introduce a tunable scaling parameter $\omega>0$ to the noise injection which controls the regularization of the DD3IP methods by scaling the values of $A^*y$ relative to the noise variance of $\epsilon$. Our reconstruction algorithm is presented in Algorithm~\ref{table:algo_dinr}.

\begin{algorithm}[htb]
 \caption{DINR Algorithm}
 \label{table:algo_dinr}
 \begin{algorithmic}[1]
     \State $S \gets $ 3D coordinate grid
     \State $A \gets $ sparse-view parallel beam projector
     \State $y \gets $ sparse-view CT projection
     \State $\phi_T \gets \arg\min_{\phi}\left\{\mathcal{L}_{\phi}(S, y, -, \rho=0)\right\} $
     \State $\theta_T \gets$ pretrained diffusion weights
     \State $x_{T} \gets \sqrt{\alpha_{T}} A^*y + \sqrt{1-\alpha_{T}}\epsilon*w$
     \For{$t = T$ to 1 timesteps:} 
        \State $\theta_{t-1} \gets \arg\min_{\theta}\left\{\text{MSE}\left(A D_\theta(x_{t}|y), y\right)\right\} $
        \State $\hat{x}_{t} \gets D_{\theta_{t-1}} \left(x_{t}|y\right)$
        \State $\phi_{t-1} \gets \arg\min_{\phi}\left\{\mathcal{L}_{\phi}(S, y, \hat{x}_{t}, \rho)\right\} $
        \If{$t==1$}
            \State $x_{t-1} \gets F_{\phi_{t-1}}(S,A^*y) $
        \Else
            \State $x_{t-1} \gets DDIM_{\theta_{t-1}}\left(F_{\phi_{t-1}}(S, A^*y), \eta\right) $
        \EndIf
     \EndFor
     \State Return $x_0$
 \end{algorithmic} 
\end{algorithm}



\vspace{-0.3cm}

\section{Results}
\label{sec:result}

In this section, we present data reconstruction results from both simulated and real observations of concrete cylinder cross-sections in an neutron CT scanner system and compare the results of our proposed method, DINR, against traditional FBP reconstructions as well as two state-of-the-art (SOTA) methods, DD3IP~
\cite{chung2024deep} and Siren~\cite{sitzmann2020implicit} (which we refer to as INR). 
Additionally, we present reconstructions of the real dataset made using MBIR with a finely tuned qGGMRF prior.

\begin{figure}[htb]
\begin{minipage}[b]{1.0\linewidth}
  \centering
  \centerline{\includegraphics[width=8.5cm]{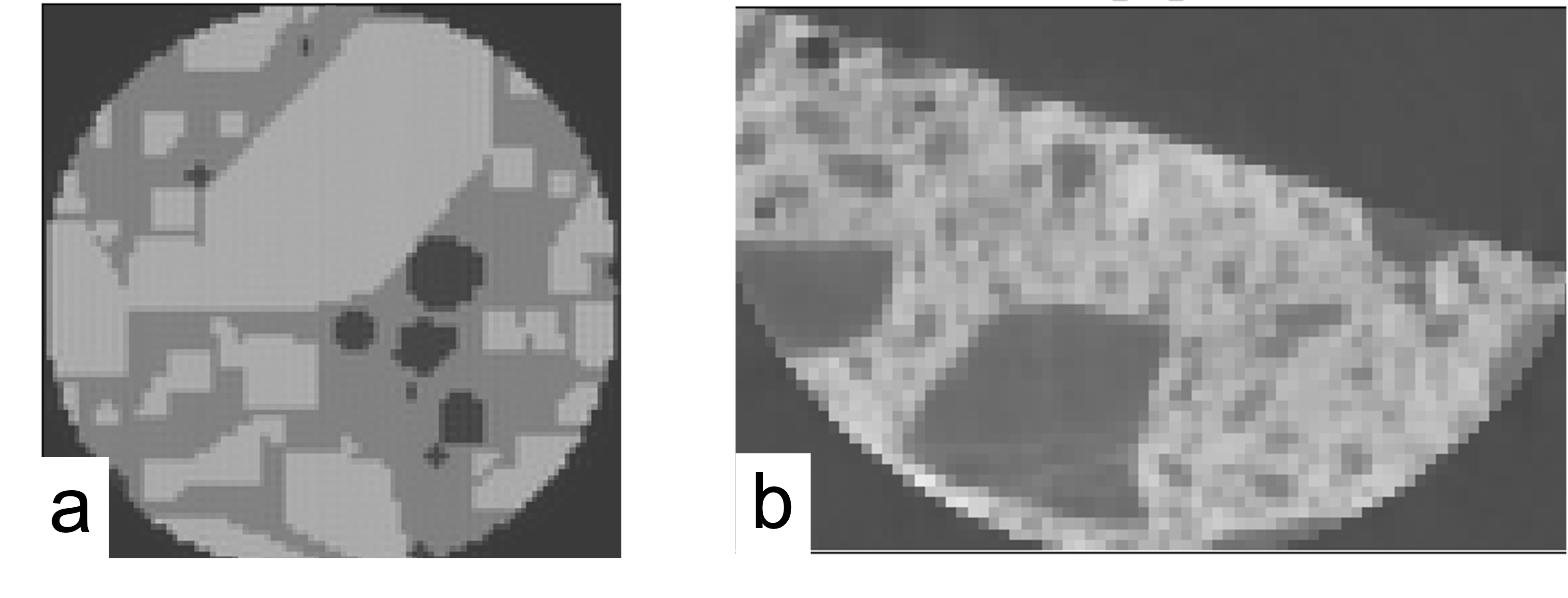}}
\end{minipage}
\caption{Reference reconstruction for comparison with images in Figures~\ref{fig:synth_results} and ~\ref{fig:real_results}, as well to calculate metrics in Tables~\ref{tab:synth_comp} and ~\ref{tab:real_comp}.}
\label{fig:refs}
\vspace{-0.5cm}
\end{figure}

\begin{figure}[htb]
\begin{minipage}[b]{1.0\linewidth}
  \centering
  \centerline{\includegraphics[width=8.5cm]{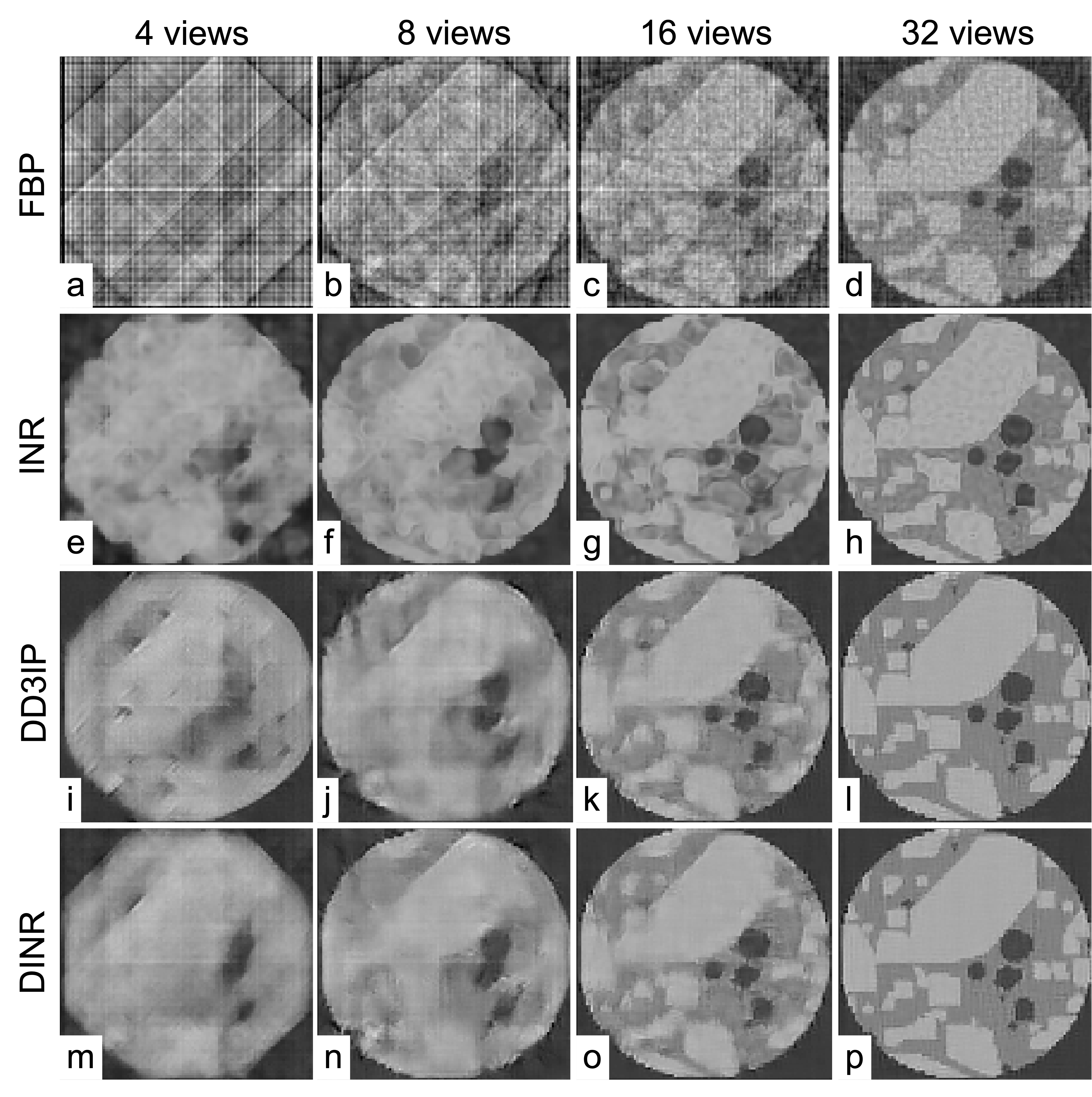}}
\end{minipage}
\caption{Reconstruction of synthetic data. The columns and rows in panels a-p are annotated to distinguish different experiments. 
DINR overall preserves the boundary and texture (pores/microstructure) very well, even for a 4-view ultra-sparse scan.}
\vspace{-0.75cm}
\label{fig:synth_results}
\end{figure}

The simulated phantom is a $2\times 256\times 256$ volume whose first slice is shown in Figure~\ref{fig:refs}(a). 
The phantom is forward projected with the tomosipo parallel beam projector at different view sparsities.
The reconstruction from the sinograms using the aforementioned methods are presented in Figure~\ref{fig:synth_results}
and corresponding performance metrics measured in PSNR and SSIM are presented in Table~\ref{tab:synth_comp}.
For the DD3IP reconstruction, we use the default parameters for learning rates, epochs, etc, provided for the UNet demo in the github associated with \cite{chung2024deep}, except the number of CG steps for the DIS is increased to 200. For both DD3IP and DINR, $\omega$ is parameter swept to maximize PSNR. For DINR, $\rho$ was chosen so that the ratio of the proximal term to the data fidelity term in $\mathcal{L}_{\phi}$ equaled $1e-5$. 

\begin{figure}[htb]
\begin{minipage}[b]{1.0\linewidth}
  \centering
  \centerline{\includegraphics[width=8.5cm]{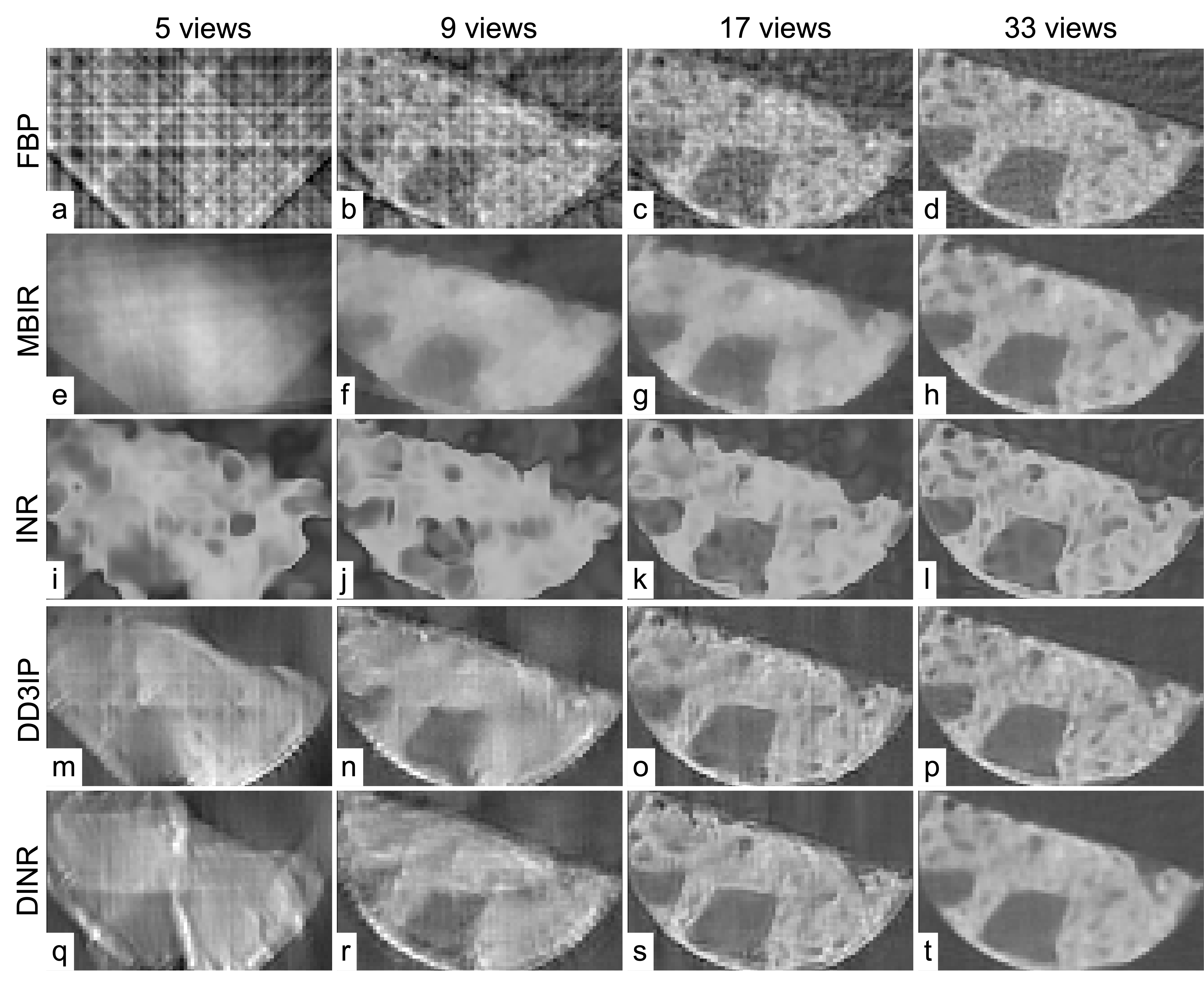}}
\end{minipage}
\caption{Reconstruction of real data. The columns and rows in panels a-t are annotated to distinguish different experiments. DINR overall preserves the boundary and texture (pores/microstructure) very well, even for a 5-view ultra-sparse scan.}
\label{fig:real_results}
\vspace{-0.25cm}
\end{figure}

The experimental sinogram is obtained from a neutron CT scanner with 1091 view angles spanning $360^\circ$, 2048 slices 
and 1048 detectors per view. 
A full-view sample is reconstructed using MBIR with a qGGMRF prior using pyMBIR software~\cite{pyMBIR} (Figure~\ref{fig:refs}(b)) and used as reference for computing metrics in Table~\ref{tab:real_comp}. 
To suppress noise and artifacts in the raw projection data, a $7 \times 7$ median filter was applied, and a regularization parameter sweep was performed to obtain the MBIR reconstruction with the highest local signal-to-noise ratio (SNR) between the foreground and background regions of the reconstructed volume.
The real sinogram (detector data) was then trimmed to 1024 detectors per view and downsampled to 256, and the central two slices were reconstructed at various view sparsities. Specifically, the original 545 view angles between $0^\circ$ and $180^\circ$ were sub-sampled by factors of 16, 32, 64, and 128, resulting in scans with 5, 9, 17, and 33 views, as summarized in Table~\ref{tab:real_comp}. Reconstructions for each experiment are shown in Figure~\ref{fig:real_results}.
For DD3IP and DINR, we set $\omega = 0.002$. Additionally, for DINR, $\rho$ was chosen such that the ratio of the proximal term to the data-fidelity term in $\mathcal{L}$ was $1\times10^{-6}$, which yielded the optimal PSNR scores.
For MBIR reconstruction of the sparse-view data, a full logarithmic sweep of the regularization parameter from $10^{-4}$ to $10^{6}$ was performed to identify the optimal value, i.e. the one yielding the best reconstruction quality compared to the ground truth, at each sparsity level.
This configuration is then used for comparison against the deep learning–based methods. While such exhaustive parameter tuning is computationally intractable for large-scale datasets, it ensures a fair and unbiased evaluation by benchmarking our method against the best achievable MBIR performance.

\begin{table}[t]
\centering
\small
\setlength{\tabcolsep}{3pt}
\renewcommand{\arraystretch}{1.2}
\caption{PSNR (dB)/SSIM comparison for different views (synthetic data). Best results are shown in bold }
\vspace{-0.1cm}
\begin{tabular}{c|p{1.5cm}p{1.5cm}p{1.6cm}p{1.5cm}}
\hline
\textbf{Views} & \textbf{FBP} & \textbf{INR} & \textbf{DD3IP} & \textbf{DINR} \\
\hline
4  & 19.31/0.08 & 14.76/0.18 & 26.17/\textbf{0.25} ($\omega = 0.02$) & \textbf{26.27}/0.24 ($\omega = 0.2$) \\
\hline 
8  & 21.67/0.18 & 28.15/0.35 & 28.37/0.34 ($\omega = 0.2$) & \textbf{28.56/0.38} ($\omega = 0.02$) \\
\hline
16 & 25.27/0.30 & 30.34/0.54 & 31.21/0.61 ($\omega = 0.02$) & \textbf{31.3/0.63} ($\omega = 0.02$) \\
\hline
32 & 29.62/0.43 & 32.85/0.66 & 32.91/0.74 ($\omega = 0.002$) & \textbf{33.43/0.76} ($\omega = 0.02$) \\
\hline
\end{tabular}
\vspace{-0.6cm}
\label{tab:synth_comp}
\end{table}



\begin{table}[t]
\centering
\small
\setlength{\tabcolsep}{3pt}
\renewcommand{\arraystretch}{1.2}
\caption{PSNR (dB)/SSIM comparison for different number of views (real data). While MBIR and DD3IP have comparable PSNR/SSIM, DINR produces higher visual quality reconstructions for ultra-space scans (Figure~\ref{fig:real_results}), highlight the needs for better quantitative metrics in future work.}
\vspace{-0.1cm}
\begin{tabular}{c|p{1.4cm}p{1.4cm}p{1.4cm}p{1.4cm}p{1.4cm}}
\hline
\textbf{Views} & \textbf{FBP} & \textbf{MBIR} & \textbf{INR} & \textbf{DD3IP} & \textbf{DINR} \\
\hline
5  & 19.9/0.1 & 21.02/0.04 & 20.18/0.03 & 20.89/\textbf{0.06} & \textbf{21.27}/0.05 \\
\hline 
9  & 22.9/0.33 & \textbf{26/0.38} & 24.08/0.27 & 25.41/0.34 & 25.22/0.35 \\
\hline
17 & 25.91/0.55 & \textbf{28.1}/0.58 & 27.3/0.54 & 28.04/\textbf{0.62} & 27.56/\textbf{0.62} \\
\hline
33 & 30.11/0.73 & 31/0.77 & 29.7/0.71 & 31.19/\textbf{0.79} & \textbf{31.37}/0.77 \\
\hline
\end{tabular}
\label{tab:real_comp}
\vspace{-0.25cm}
\end{table}


One of the key challenges in evaluating reconstruction quality for scientific imaging data lies in defining an appropriate region of interest (ROI) where an algorithm performs best. In the literature, ROI-based metrics are often computed using manually defined masks; however, this approach assumes prior knowledge of where to mask and introduces user bias, which is particularly problematic for microstructure data with irregular geometries. 
To minimize this bias, we adopt a consistent, data-driven approach. We first identify the largest region within the field of view that contains both the object and background without visible artifacts in the ground-truth reconstruction, an area of $64 \times 96$ pixels. From this region, smaller subregions of $64 \times 64$, $48 \times 48$, $32 \times 32$, $16 \times 16$, and $8 \times 8$ pixels are cropped systematically, ensuring automatic and unbiased ROI selection. Within each subregion, PSNR is computed for all reconstruction methods.

\begin{figure}[htb]
\begin{minipage}[b]{1.0\linewidth}
  \centering
  \centerline{\includegraphics[width=8.5cm]{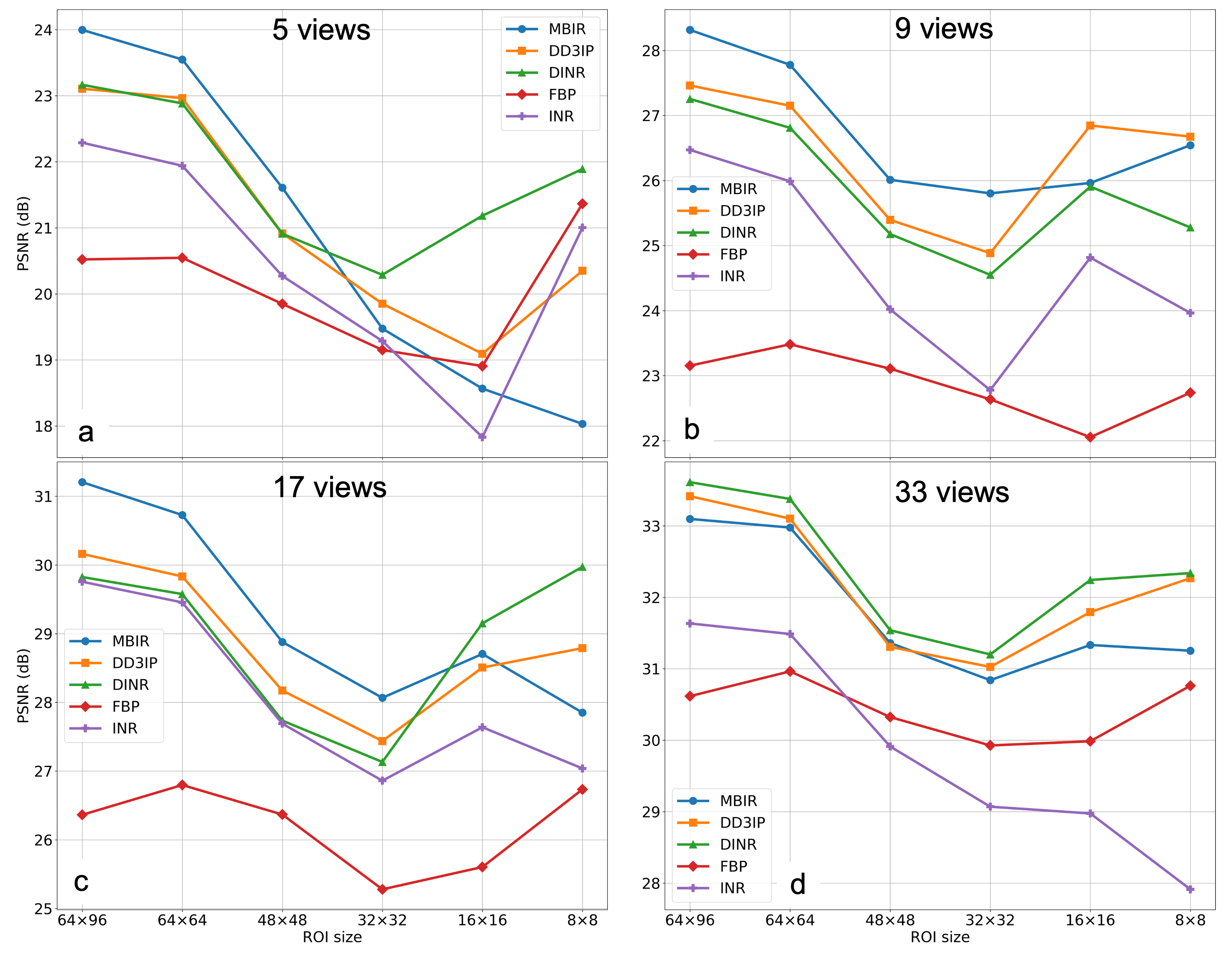}}
\end{minipage}
\caption{PSNR calculated systematically for varying ROI to mask the impact of the background and compare performance in reducing artifacts within microstructure region for a) 5 views, b) 9 views, c) 17 views, and d) 33 views.}
\label{fig:roi_metric}
\vspace{-0.5cm}
\end{figure}

As shown in Figure~\ref{fig:roi_metric}, for reconstructions with 5, 9, 17, and 33 views, the DINR method demonstrates superior performance within regions dominated by microstructural features rather than background, particularly for ROIs smaller than $48 \times 48$ pixels and most notably below $32 \times 32$ pixels. 
This indicates that DINR achieves higher fidelity within the microstructure, while inclusion of background regions reduces its relative advantage and allows other techniques to attain similar quantitative metrics. 
This trend is also qualitatively evident in the 5/9-view MBIR reconstruction results, where the overall MBIR reconstruction quality in Figure~\ref{fig:real_results}(e)-(f) appears worse than DD3IP and DINR, yet the MBIR method yields higher PSNR in ROIs that include a larger portion of the background.

This analysis is conceptually analogous to the \textit{SNR growth-curve} or \textit{task-based detectability} methods used in CT image-quality evaluation, where SNR is analyzed as a function of ROI size to assess spatial resolution and local contrast fidelity. Future work will focus on developing more discriminative metrics based on segmentation of particles and boundaries within the microstructure to further refine quantitative evaluation.

\vspace{-0.4cm}
\section{Conclusion}
\label{sec:conlcusion}
We proposed DINR for sparse-view CT reconstruction and demonstrated its capabilities over a state of-the-art diffusion inverse solver and INR. In particular, we demonstrated results for synthetic and real neutron CT of microstructure data, and observed that even with a small number of views (min. 5 in our test data), our proposed DINR produces higher quality reconstructions than SOTA approaches. 
However, the efficacy of our method for handling view sparsity can be extended to other CT beam sources such as X-ray, or electron CT. 
In future work, we intend to conduct a comprehensive ablation study to evaluate the contribution of the FBP input to both INR and DINR reconstructions. Furthermore, we anticipate that a more exhaustive parameter search will establish that DINR can outperform MBIR across a broader range of imaging conditions. Finally, we plan to extend our software framework to support multi-GPU implementations, enabling the reconstruction of larger volumetric datasets and facilitating the application of DINR to more advanced CT acquisition geometries, including cone-beam and helical CT and extend our work to other CT applications that require sparse view sampling.

\vspace{-0.4cm}


\bibliographystyle{IEEEbib}
\bibliography{strings,refs}

\end{document}